\documentclass[aps,prc,a4paper,twocolumn,nofootinbib]{revtex4}
\usepackage{graphicx}
\usepackage{bm}
\usepackage{amsmath}
\usepackage{amsfonts}

\usepackage{mathptmx, courier, pifont}
\usepackage[scaled=0.92]{helvet}
\usepackage[T1]{fontenc}
\usepackage{textcomp}

\begin{document}
\title{Grassmann integral and Balian-Br\'ezin decomposition
in Hartree-Fock-Bogoliubov matrix elements}

\author{Takahiro Mizusaki$^1$, Makito Oi$^1$, Fang-Qi Chen$^2$, Yang Sun$^{2,3}$}
\affiliation{$^1$ Institute of Natural Sciences, Senshu University,
3-8-1 Kanda-Jinbocho, Chiyoda-ku, Tokyo 101-8425, Japan\\
$^2$Department of Physics and Astronomy, Shanghai Jiao Tong University, Shanghai
200240, People's Republic of China\\
$^3$Institute of Modern Physics, Chinese Academy of Sciences,
Lanzhou 730000, People's Republic of China}


\begin{abstract}
We present a new formula to calculate matrix elements 
of a general unitary operator with respect to Hartree-Fock-Bogoliubov states
allowing multiple quasi-particle excitations. 
The Balian-Br\'ezin decomposition of the unitary operator 
(Il Nuovo Cimento B 64, 37 (1969)) is employed in the derivation.
We found that this decomposition is extremely suitable for 
an application of Fermion coherent state and Grassmann integrals
in the quasi-particle basis.
The resultant formula is compactly expressed in terms of the Pfaffian, 
and shows the similar bipartite structure to the formula that 
we have previously derived in the bare-particles basis 
(Phys. Lett. B 707, 305 (2012)).

\end{abstract}

\maketitle
\section{Introduction}
In nuclear many-body physics, evaluations of matrix elements of
many-body operators have been a major obstacle to implementations of
sophisticated methods and theories beyond the mean-field approximation.
Nuclear physicists have put effort into
finding convenient formulae \cite{OY66,HI79,NW83} to calculate matrix elements (and overlaps) 
with respect to Hartree-Fock-Bogoliubov (HFB) states.
But difficulties remained in the obtained formulae.
Although there were some numerical attempts to circumvent 
difficulties associated with the known formulae \cite{HHR80,OT05}, 
there had not been any significant progress for decades 
in an analytical attempt to make a breakthrough.
Recently such a breakthrough was achieved  by Robledo 
who was successful in deriving a new formula in terms of the Pfaffian \cite{Rob09} 
with  Fermion coherent states and Grassmann integral \cite{NO85}. 
After his pioneering work, 
many studies followed by exploiting 
these mathematical tools in the HFB matrix elements\cite{BR12,OM11,MO12,AB12}.

The latest focus in this research field is to find a formula
to evaluate HFB matrix elements with multiple quasi-particle excitations 
\cite{OM11, MO12, AB12}
$\langle \phi'|c_{\nu'_1} \cdots c_{\nu'_{n'}} c^\dag_{\nu_1}
\cdots c^\dag_{\nu_n} | \phi\rangle$, 
where  $| \phi\rangle $ and  $| \phi'\rangle $ are different HFB states.
The creation and annihilation operators for bare particles
are denoted by $c^{\dag}$ and $c$ respectively, hence $c_\nu|0\rangle=0$.
$|0\rangle$ stands for the true vacuum.
These matrix elements have been evaluated conventionally 
by the generalized Wick's theorem.
Recently, alternative approaches \cite{OM11,MO12,AB12} 
were obtained by means of  Fermion coherent states and  Grassmann integral.
The resultant formulae are expressed in terms of the Pfaffian, 
which can describe the matrix elements in a more compact manner
than those obtained by the generalized Wick's theorem.
The new formulae overcome a combinatorial complexity associated with
practical applications of the generalized Wick's theorem.
(It is worth mentioning that there was an attempt to derive 
a compact formula before the publication of Robledo's Pfaffian formula.
Ref.\cite{Rob07} adapts Gaudin's theorem in the
finite-temperature formalism so as to derive an equivalent formula, but
it is not expressed in terms of the Pfaffian. )

In this Letter, we would like to present another formula that evaluates
matrix elements of a unitary operator sandwiched by 
arbitrary HFB states with multiple quasi-particle excitations,
\begin{equation}
\langle \phi|a_{\nu'_1} \cdots a_{\nu'_{n'}} [\theta] a^\dag_{\nu_1}
\cdots a^\dag_{\nu_n} | \phi\rangle.
\label{overlap}
\end{equation}
The quasi-particle basis $(a,a^\dag)$ is obtained through
a canonical transformation (called the Bogoliubov transformation)
of the bare-particle basis $(c,c^{\dag})$, 
\begin{equation}
a_{\nu}^{\dagger} = \sum_{i=1}^M \left( U_{i,\nu}c{_i}^{\dag} +
V_{i,\nu}c_i \right), \label{UV}
\end{equation}
$M$ is the dimension of the single-particle model space, which is
taken to be an even integer.
Coefficients $U$ and $V$ in the expression above
define the Bogoliubov transformation. 
HFB state $| \phi\rangle $ is also obtained through the Bogoliubov
transformation applied to the bare-particle vacuum $|0\rangle$,
hence $|\phi\rangle$ corresponds to the vacuum for the quasi-particles, 
or $a_\nu | \phi\rangle =0$.
Indices $\nu_1,\cdots, \nu_n$ and  $\nu'_1,\cdots, \nu'_{n'}$ ,
attached to the creation and annihilation operators,
specify quantum states in the quasi-particle basis.
The symbol $[\theta]$ stands for a unitary operator
\begin{equation}
[\theta]\equiv\frac{e^{-i\theta \hat{S}}}{\langle\phi|
e^{-i\theta\hat S} | \phi\rangle},
\label{Sope}
\end{equation}
where $\hat{S}$ is a one-body operator expressed in the quasi-particle basis
\cite{RS80}, and $\theta$ is a parameter to specify an element
in a group produced by the generator $\hat{S}$.
Symbol $\left[\theta\right]$ for unitary operators 
is indebted to Ref.\cite{HI79}.

As explained in Ref.\cite{HI79}, such matrix elements of unitary operators 
shown in Eq.(\ref{overlap}) are essential ingredients  in the beyond-mean-field
theories, such as quantum number projection.
In the case of angular momentum projection, the parameter $\theta$
corresponds to the Euler angles 
and $\hat{S}$ to the angular momentum operator $\hat{J}$.

In order to derive a formula for this matrix element Eq.(\ref{overlap}), 
we will apply the Fermion coherent state and Grassmann integral.
As elucidated in previous studies, these two mathematical entities 
show a close affinity with the Pfaffian, and 
they can simplify calculations involving many anti-commuting
operators to a great extent.
It should be noted here that Hara  and Iwasaki previously 
investigated the mathematical strucutre of the matrix elements 
Eq.(\ref{overlap}) \cite{HI79}, 
in connection to the Projected Shell Model (PSM) \cite{HS95}.
PSM is basically a configuration mixing method 
with multi quasi-particle states based on the HFB theory.
In PSM, configuration mixing is carried out with 
quantum-number projected HFB states of multiple quasi-particle excitations.
Hara and Iwasaki derived a formula for the matrix elements 
Eq.(\ref{overlap})  with the help of a theorem presented 
by Balian and Br\'ezin \cite{BB69}.
However, their formula suffers from the problem of combinatorial complexity,
originating from the generalized Wick's theorem.
According to the theorem, the matrix elements Eq.(\ref{overlap}) 
involving $n$ and $n'$ quasi-particle excitations 
contain $(n+n'-1)!!$ terms. 
In practice, the number of terms becomes so large
that it is difficult to write down matrix elements explicitly 
with the Hara-Iwasaki formula
for more than four quasi-particle HFB states.

\section{The Balian-Br\'ezin decomposition}

Following Ref. \cite{BB69}, a unitary operator 
$[\theta]$ in Eq.(\ref{Sope}) can be expressed 
as a product of three operators in the quasi-particle basis,
\begin{equation}
[\theta] = e^{\hat{B}(\theta)}e^{\hat{C}(\theta)}
e^{\hat{A}(\theta)}, \label{abc}
\end{equation}
with
\begin{eqnarray} \label{ABCmatrix}
\hat{A}(\theta)&=&\sum \frac{1}{2}A(\theta)_{\nu',\nu}a_{\nu} a_{\nu'} \nonumber \\
\hat{B}(\theta)&=&\sum \frac{1}{2}B(\theta)_{\nu',\nu}a^{\dagger}_{\nu} a^{\dagger}_{\nu'} \\
\hat{C}(\theta)&=&\sum (ln C(\theta))_{\nu,\nu'}a^{\dagger}_{\nu}
a_{\nu'} \nonumber .
\end{eqnarray}
We call Eq.(\ref{abc}) the Balian-Br\'ezin decomposition.
Matrices $A(\theta)$, $B(\theta)$ and $C(\theta)$ in Eq.(\ref{ABCmatrix})
correspond to contractions and can be written with the help of
the Bogoliubov transformation matrices \cite{HI79}
\begin{eqnarray} \label{ABC3}
A_{\nu,\nu'}(\theta)&\equiv & \langle [\theta] a^{\dagger}_{\nu} a^{\dagger}_{\nu'} \rangle = \left(V^{*}(\theta)U^{-1}(\theta)\right)_{\nu,\nu'} \nonumber \\
B_{\nu,\nu'}(\theta)&\equiv & \langle  a_{\nu} a_{\nu'} [\theta] \rangle = 
\left(U^{-1}(\theta)V(\theta)\right)_{\nu,\nu'} \\
C_{\nu,\nu'}(\theta)&\equiv & \langle  a_{\nu} [\theta]
a^{\dagger}_{\nu'} \rangle = \left(U^{-1}(\theta)\right)_{\nu,\nu'} \nonumber .
\end{eqnarray}

By inserting Eq.(\ref{abc}) into the matrix elements Eq.(\ref{overlap}),
we have
\begin{equation}
 \mathcal{M}_I=\langle \phi|a_{\nu'_1} \cdots a_{\nu'_{n'}} e^{\hat{B}(\theta)}e^{\hat{C}(\theta)}
e^{\hat{A}(\theta)} a^\dag_{\nu_1}
\cdots a^\dag_{\nu_n} | \phi\rangle.
\label{overlap-abc}
\end{equation}
Hereafter, we omit symbol $(\theta)$ for the sake of brevity. 
For subsequent conveniences, 
we introduce a shorthand notation $J$ for the indices of quasi-particle 
operators,
 as $J=\{ {\nu_1} \cdots \nu_{n} \}$,
 $J'=\{ {\nu'_1} \cdots \nu'_{n'} \}$, 
($ {\nu_1} <  \cdots < \nu_{n}  $ and $ {\nu'_1} <  \cdots < \nu'_{n'}  $).
These indices $J$ and $J'$ are subsets of a set $[M]=\{1,2,\cdots, M\}$,
in which $M$ represents the number of elements in $[M]$
and corresponds to the dimension of the single-particle model space. 
Index $I$ in Eq.(\ref{overlap-abc}) is defined as 
a set $I=\{  {\nu'_1} \cdots \nu'_{n'}, {\nu_1}+M \cdots \nu_{n}+M \}$
and corresponds to a subset of  $[2M]=\{1,2,\cdots, 2M\}$. 
With these notations, 
the matrix elements Eq.(\ref{overlap-abc}) are expressed as 
\begin{equation}
 \mathcal{M}_I=\langle \phi|(a \cdots a)_{\overrightarrow{J'}} e^{\hat{B}}e^{\hat{C}}
e^{\hat{A}} (a^\dag \cdots a^{\dag})_{\overrightarrow{J}} | \phi\rangle
\label{overlap-abc-short}
\end{equation}
where $(a \cdots a)_{\overrightarrow{J'}}$ and  $(a^\dag \cdots a^{\dag})_{\overrightarrow{J}}$ stand for
$a_{\nu'_1} \cdots a_{\nu'_{n'}}$ 
and $a^\dag_{\nu_1}\cdots a^\dag_{\nu_n}$, 
respectively.

When the order of a product is completely reversed,
such an order is denoted as $\overleftarrow{J}$.
The relation between $\overleftarrow{J}$ and $\overrightarrow{J}$ is given
as
\begin{equation}
(a \cdots a)_{\overrightarrow{J}} =(a \cdots a)_{\overleftarrow{J}}(-)^{\frac{1}{2}n(n-1)},
\label{reverse}
\end{equation}
where  $(a \cdots a)_{\overleftarrow{J}}=a_{\nu_{n}} \cdots  a_{\nu_1}  $. 
Note that an additional phase emerges in the right-hand side of the 
above equation due to anti-commutation.

\section{Fermion Coherent state and Grassmann integral }

In the present paper, 
we exclusively rely on Grassmann numbers  $\xi^*$ and $\xi$.
They satisfy the anti-commutation rules,
\begin{eqnarray}
\xi_\nu \xi_{\nu'} + \xi_{\nu'} \xi_\nu &=& 0,\\
\xi_\nu^*\xi_{\nu'}^* + \xi_{\nu'}^*\xi_\nu^* &=& 0, \\
\xi_\nu \xi_{\nu'}^* + \xi_{\nu'}^*\xi_\nu &=& 0, 
\end{eqnarray}
where indices $\nu,\nu'$ run from $1$ to $M$ ($1,\cdots,M$).
With these Grassmann numbers, Fermion coherent states \cite{NO85}
in the quasi-particle basis are defined as 
\begin{equation}
  |\bm{\xi}\rangle = e^{-\sum_{\nu}\xi_{\nu}a_{\nu}^{\dag}}|\phi \rangle,
  \label{FCS}
\end{equation}
where the HFB state is normalized $\langle \phi|\phi \rangle=1$.
This definition is slightly different from the one 
introduced in Ref.\cite{MO12}, where the operator and the vacuum
are replaced as $c_i \rightarrow a_\nu$ and $|0\rangle\rightarrow |\phi\rangle$.
By definition,  Fermion coherent states are eigenstates of the annihilation
operator,
\begin{equation}
  a_\nu|\bm{\xi}\rangle = \xi_\nu|\bm{\xi}\rangle.
  \label{c-map}
\end{equation}
The adjoint variable $\xi_i^*$ is also introduced in the eigenvalue equation,
\begin{equation}
  \langle \bm{\xi} |a_\nu^{\dagger} = \langle \bm{\xi} |\xi_\nu^*.
  \label{cdag-map}
\end{equation}
The overlap between the HFB vacuum and the Fermion coherent state
is  $\langle \phi|\xi\rangle=1$.
The closure relation \cite{NO85} is expressed  as 
\begin{equation}
\int \mathcal{D}(\xi^*, \xi) e^{-\Sigma_\nu \xi_\nu^* \xi_\nu}
|\bm{\xi}\rangle  \langle \bm{\xi}|=1,
\label{complete}
\end{equation}
where $\mathcal{D}(\xi^*, \xi)= \prod_{\alpha}d\xi_\alpha^* d\xi_\alpha  $.
Differential elements $d\xi$ and $d\xi^*$ are anti-commuting. 
Although  this ordering for $\mathcal{D}(\xi^*, \xi)$ given 
in the above closure relation
is widely employed, we use other ordering for the differential elements
in the present study, which is
\begin{equation}
\mathcal{D}(\xi^*, \xi)=d\xi^{*}_{\overrightarrow{[M]}} d\xi_{\overleftarrow{[M]}}
=d\xi_{\overleftarrow{[M]}} d\xi^{*}_{\overrightarrow{[M]}},
\label{two-derivative}
\end{equation}
where $(-)^M =1$ is used because $M$ is even.  
``Half products'' in the above expression are defined as
\begin{eqnarray}
d\xi_{\overrightarrow{[M]}}&=&d \xi_1 \cdots d \xi_M, \nonumber \\
d\xi_{\overleftarrow{[M]}}&=&d \xi_M \cdots d \xi_1,
\end{eqnarray}
where $[M]=\{1,2,\cdots, M\}$.
This definition of the ordering is also applied to Grassmann numbers $\xi^*$.

For the sake of convenience in later discussions,
we need to define several other orderings for ``partial products''
of the differential elements.
Firstly, we define products relevant to the indices 
$J=\{\nu_1,\cdots, \nu_n\}$ appearing
in the matrix element Eq.(\ref{overlap-abc-short}),
\begin{eqnarray}
d\xi_{\overrightarrow{J}}&=&d \xi_{\nu_1} \cdots d \xi_{\nu_n} \nonumber \\
d\xi_{\overleftarrow{J}}&=&d \xi_{\nu_n} \cdots d \xi_{\nu_1}.
\end{eqnarray}
This partial product was originally introduced in Ref.\cite{AB12}.
The second type of partial products is defined for
the complement set $\overline{J}=[M]-J$,
\begin{eqnarray}
d\xi_{\overrightarrow{\overline{J}}}&=&
d\xi_1d\xi_2\cdots
\hat{d\xi}_{\nu_1}\cdots \hat{d \xi}_{\nu_n}\cdots d\xi_{M-1}d\xi_M 
\nonumber \\
d\xi_{\overleftarrow{\overline{J}}}&=&
d\xi_Md\xi_{M-1}\cdots \hat{d \xi}_{\nu_n} \cdots \hat{d \xi}_{\nu_1}\cdots
d\xi_2d\xi_1,
\end{eqnarray}
where symbol $\hat{d\xi}_i$ here means a removal of $d\xi_i$
from the product $d\xi_{\overrightarrow{[M]}}$ (or $d\xi_{\overleftarrow{[M]}}$).
For an example,  $d\xi_1\hat{d\xi}_2 d\xi_3d\xi_4=d\xi_1d\xi_3d\xi_4$.
Combining these two types of partial products, we would like to rewrite
a half product $d\xi_{\overleftarrow{[M]}}$ 
(as well as $d\xi_{\overrightarrow{[M]}}$) as,
\begin{eqnarray}
d\xi_{\overleftarrow{[M]}}&=&d\xi_{\overleftarrow{\overline{J}}}d\xi_{\overleftarrow{J}}(-)^{|J|-\frac{n(n+1)}{2}} \nonumber \\
d\xi_{\overrightarrow{[M]}}&=&d\xi_{\overrightarrow{\overline{J}}}d\xi_{\overrightarrow{J}}(-)^{|J|+\frac{n(n-1)}{2}}
\label{split_diff}
\end{eqnarray}
where $|J|\equiv\sum_k \nu_k$. 
The latter relation can be obtained by reversing the arrow directions
of $[M]$ and $J$ in the former relation.
The relevant phase emerging due to the reordering
can be calculated by referring to Eq.(\ref{reverse}).
Noting that the complement set of $\overline{J}$ comes back to 
$J(=\overline{\overline{J}})$,
it is also possible to obtain the following ordering from Eq.(\ref{split_diff}),
\begin{eqnarray}
d\xi_{\overleftarrow{[M]}}&=&d\xi_{\overleftarrow{J}}d\xi_{\overleftarrow{\overline{J}}}(-)^{|\overline{J}|-\frac{\overline{n}(\overline{n}+1)}{2}} \nonumber \\
d\xi_{\overrightarrow{[M]}}&=&d\xi_{\overrightarrow{J}}d\xi_{\overrightarrow{\overline{J}}}(-)^{|\overline{J}|+\frac{\overline{n}(\overline{n}-1)}{2}}.
\label{split_diff_bar}
\end{eqnarray}
The number of the elements of  $\overline{J}$ is denoted as $\overline{n}=M-n$.
The relations Eqs.(\ref{split_diff},\ref{split_diff_bar}) also hold for $\xi^*$.

In summary, 
by these relations Eqs.(\ref{split_diff},\ref{split_diff_bar}) 
obtained above,
the differential elements in total can be expressed as
\begin{eqnarray}
d\xi^*_{\overrightarrow{[M]}}d\xi_{\overleftarrow{[M]}}&=&d\xi^*_{\overrightarrow{J}}d\xi^*_{\overrightarrow{\overline{J}}}
d\xi_{\overleftarrow{\overline{J}}}d\xi_{\overleftarrow{J}}  \nonumber \\
d\xi_{\overleftarrow{[M]}}d\xi^*_{\overrightarrow{[M]}}&=&d\xi_{\overleftarrow{J}}d\xi_{\overleftarrow{\overline{J}}}
d\xi^*_{\overrightarrow{\overline{J}}}d\xi^*_{\overrightarrow{J}},
\label{xixi}
\end{eqnarray}
where we used $|J|+|\overline{J}|=\frac{1}{2}M(M+1)$.

\section{Matrix element in terms of Grassmann integral}
Now, let us introduce 
two sets of Grassmann numbers 
\{ $\xi^*$, $\xi$ \} and  \{ $\eta^*$, $\eta$ \}.  
These Grassmann numbers are anti-commuting each other.
We then insert two kinds of the closure relations Eq.(\ref{complete}) 
for $\xi$ and $\eta$ to the matrix element Eq.(\ref{overlap-abc}).
The closure relation for $\xi$ ($\eta$) is inserted
in the position between $a_{\nu'_{n'}}$ and $e^{\hat{B}}$ 
($e^{\hat{A}}$ and  $a^\dag_{\nu_1}$).
Thus, the matrix element $ \mathcal{M}_I$  becomes 
\begin{eqnarray}
\mathcal{M}_I&=&
\int \mathcal{D}(\xi^*, \xi)  \langle \phi|(a \cdots a)_{\overrightarrow{J'}}|\xi\rangle e^{-\Sigma_\nu \xi_\nu^* \xi_\nu}   \nonumber \\
&& \int  \mathcal{D}(\eta^*, \eta)  \langle \xi |e^{\hat{B}}e^{\hat{C}}e^{\hat{A}}|\eta\rangle  e^{-\Sigma_\nu \eta_\nu^* \eta_\nu} \nonumber \\  
&& \langle \eta| (a^{\dag}\cdots a^\dag)_{\overrightarrow{J}} | \phi\rangle.
\label{Mabc}
\end{eqnarray}
At this stage, all the operators are replaced with Grassmann numbers.
First of all, the following relations hold:
\begin{equation}
 \langle \phi| (a \cdots a)_{\overrightarrow{J'}}|\xi\rangle=(\xi \cdots \xi)_{\overrightarrow{J'}}
 \label{firstterm}
\end{equation} 
and 
\begin{equation}
\langle \eta | (a^\dag \cdots a^\dag )_{\overrightarrow{J}} | \phi\rangle
=(\eta^{*} \cdots \eta^*)_{\overrightarrow{J}}.
 \label{lastterm}
\end{equation} 
Here,  we used $\langle \phi|\xi\rangle=1$ and  $\langle \phi|\eta\rangle=1$,
as well as Eqs.(\ref{c-map},\ref{cdag-map}).
We also introduced shorthand notations for Grassmann numbers $\xi$ and $\eta$ 
in the above relation, as
$(\xi \cdots \xi)_{\overrightarrow{J'}}=\xi_{\nu'_1} \cdots \xi_{\nu'_{n'}}$  and $(\eta^{*} \cdots \eta^*)_{\overrightarrow{J}}=\eta^{*}_{\nu_1} \cdots \eta^*_{\nu_n}$. 

Secondly, we deal with a product of exponential operators
$ e^{\hat{B}}e^{\hat{C}} e^{\hat{A}}$.
Because $ \hat{A} $ and $\hat{B}$ 
do not contain any $a^{\dag}a$ terms,
the expectation value of $ e^{\hat{B}}e^{\hat{C}} e^{\hat{A}}$ is transformed as 
\begin{equation}
\langle \xi |e^{\hat{B}} e^{\hat{C}}e^{\hat{A}}|\eta\rangle
=
 e^{\sum \frac{1}{2}B_{\nu',\nu}\xi^{*}_{\nu} \xi^{*}_{\nu'}}
 \langle \xi |e^{\hat{C}}|\eta\rangle
 e^{\sum \frac{1}{2}A_{\nu',\nu}\eta_{\nu} \eta_{\nu'}}.
 \label{middleterm}
\end{equation} 
To evaluate $ \langle \xi |e^{\hat{C}}|\eta\rangle$,
we use a following relation  Eq.(\ref{grassmann-a+a}) as
\begin{equation}
e^{\hat{C}}|\eta\rangle=|\sum C_{\nu, \nu'} \eta_{\nu'}\rangle,
\end{equation}
which is derived in Appendix.
Thus, $\langle \xi |e^{\hat{C}}|\eta\rangle$ term becomes 
\begin{equation}
\langle \xi |e^{\hat{C}}|\eta\rangle=\langle \xi |C \eta\rangle=e^{\sum \xi^*_\nu C_{\nu, \nu'} \eta_{\nu'}},
\label{xce}
\end{equation}
where we use Eq.(\ref{grassmann-overlap}) in Appendix.

Together with Eqs.(\ref{two-derivative},\ref{firstterm},\ref{lastterm},\ref{middleterm},\ref{xce}),
Eq.(\ref{Mabc}) is represented in terms of the Grassmann integrals,
\begin{eqnarray}
\mathcal{M}_I&=&\int d\xi^{*}_{\overrightarrow{[M]}} d\xi_{\overleftarrow{[M]}}
 (\xi \cdots \xi)_{\overrightarrow{J'}} e^{-\Sigma_\nu \xi_\nu^* \xi_\nu}  \nonumber \\
& &\int  d\eta_{\overrightarrow{[M]}}  d \eta^* _{\overleftarrow{[M]}}
(\eta^* \cdots \eta^*)_{\overrightarrow{J}}  e^{-\Sigma_\nu \eta_\nu^* \eta_\nu} 
B(\xi^*,\eta),
\label{Mabc2}
\end{eqnarray}
where we use Eq.(\ref{two-derivative}) for the differential elements
and  
\begin{equation}
B(\xi^*,\eta)\equiv e^{\sum \frac{1}{2}B_{\nu',\nu}\xi^*_{\nu} \xi^*_{\nu'}}
e^{\sum \xi^*_\nu C_{\nu, \nu'} \eta_{\nu'}}
e^{\sum \frac{1}{2}A_{\nu',\nu}\eta_{\nu} \eta_{\nu'}}.
\end{equation}
We should note that $B(\xi^*,\eta)$ is a function of only $\xi^*$ and $\eta$, 
but not of $\xi$ and $\eta^*$.

\section{Evaluation of Grassmann integral}
Now we carry out these Grassmann integrals in Eq.(\ref{Mabc2}). 
Part of Eq.(\ref{Mabc2}) concerning Grassmann integral 
over $ \eta$ and $ \eta^*$  is rewritten as 
\begin{eqnarray}
&& \int  d\eta_{\overrightarrow{[M]}}  d \eta^* _{\overleftarrow{[M]}}
(\eta^* \cdots \eta^*)_{\overrightarrow{J}}  e^{-\Sigma_\nu \eta_\nu^* \eta_\nu} 
B(\xi^*,\eta) \nonumber \\
&=&\int 
d\eta_{\overrightarrow{J}}d\eta_{\overrightarrow{\overline{J}}}  B(\xi^*,\eta)
\int d\eta^*_{\overleftarrow{\overline{J}}}d\eta^*_{\overleftarrow{J}}
(\eta^* \cdots \eta^*)_{\overrightarrow{J}}  e^{-\Sigma_\nu \eta_\nu^* \eta_\nu},\nonumber \\ 
\label{etas}
\end{eqnarray}
where we use  Eq.(\ref{xixi}). 
The integral over $\eta^*_\nu (\nu \in J)$ gives rise to unity because
quadratic forms of Grassmann variables vanish by definition ($\xi_\nu^2=0$).
As a result,
\begin{equation}
\int d\eta^*_{\overleftarrow{J}}
(\eta^* \cdots \eta^*)_{\overrightarrow{J}}  e^{-\Sigma_\nu \eta_\nu^* \eta_\nu} =1.
\end{equation}
The integral over $\eta^*_\nu (\nu \in \overline{J})$ 
gives rise to a product of $\eta$'s as 
\begin{equation} 
\int d\eta^*_{\overleftarrow{\overline{J}}} e^{-\sum_{\nu \in  \overline{J}} \eta_\nu^* \eta_\nu}=
(-)^{\overline{n}}(\eta \cdots \eta)_{\overleftarrow{\overline{J}}},
\end{equation}
where $\overline{n}=M-n$. Then the integral over $\eta_\nu (\nu \in \overline{J})$
in the Eq.(\ref{etas}) can be carried out as
\begin{equation}
\int d\eta_{\overrightarrow{J}}d\eta_{\overrightarrow{\overline{J}}} 
(-)^{\overline{n}}(\eta \cdots \eta)_{\overleftarrow{\overline{J}}}
 B(\xi^*,\eta) 
=(-)^{\overline{n}} \int d\eta_{\overrightarrow{J} } B(\xi^*,\{\eta\}_J),
\end{equation}
where $\{\eta\}_J$ stands for the Grassmann variables $ \eta_{\nu_1}, \cdots, \eta_{\nu_{n}}$. 
After the integrals over $\eta_\nu (\nu \in \overline{J})$ and 
 $\eta^*_\nu$  are performed,
  the integral over $\eta_\nu (\nu \in J)$  remains. 
Then, Eq.(\ref{Mabc2}) is represented by the integrals
over $\xi$, $\xi^*$ and  $\eta_\nu (\nu \in J)$  as 
\begin{eqnarray}
\mathcal{M}_I&=&(-)^{\overline{n}}\int d\xi^{*}_{\overrightarrow{[M]}} d\xi_{\overleftarrow{[M]}}
 (\xi \cdots \xi)_{\overrightarrow{J'}} e^{-\Sigma_\nu \xi_\nu^* \xi_\nu} 
  \int d\eta_{\overrightarrow{J} } B(\xi^*,\{\eta\}_J)  \nonumber \\
&=&(-)^{\overline{n}} \int d\xi^*_{\overrightarrow{J'}}d\xi^*_{\overrightarrow{\overline{J'}}}
d\xi_{\overleftarrow{\overline{J'}}}d\xi_{\overleftarrow{J'}}   
 (\xi \cdots \xi)_{\overrightarrow{J'}} e^{-\Sigma_\nu \xi_\nu^* \xi_\nu} \nonumber \\ 
& &  \int d\eta_{\overrightarrow{J} } B(\xi^*,\{\eta\}_J).
\end{eqnarray}
In the same way, the integral over $\xi$  in the above integrals is carried out as
\begin{equation}
\int d\xi_{\overleftarrow{\overline{J'}}}d\xi_{\overleftarrow{J'}}   
 (\xi \cdots \xi)_{\overrightarrow{J'}} e^{-\Sigma_\nu \xi_\nu^* \xi_\nu} 
 = (\xi^* \cdots \xi^*)_{\overleftarrow{\overline{J'}}}.
\end{equation} 
Thus,

\begin{eqnarray}
\mathcal{M}_I&=& (-)^{\overline{n}}\int d\xi^*_{\overrightarrow{J'}}d\xi^*_{\overrightarrow{\overline{J'}}}   
 (\xi^* \cdots \xi^*)_{\overleftarrow{\overline{J'}}}
  \int d\eta_{\overrightarrow{J} } B(\xi^*,\{\eta\}_J) \nonumber  \\
  &=&(-)^{\overline{n}} \int d\xi^*_{\overrightarrow{J'}} d\eta_{\overrightarrow{J} } B({\{\xi^*\}}_{J'},\{\eta\}_J)  \nonumber  \\
  &=&(-)^{\sigma} \int d\eta_{\overleftarrow{J}} d\xi^*_{\overleftarrow{J'}}   B({\{\xi^*\}}_{J'},\{\eta\}_J),
\end{eqnarray}
where $\{\xi^* \}_J$ stands for the Grassmann variables $ \xi^*_{\nu_1}, \cdots, \xi^*_{\nu_{n}}$.
In the last line, an order of differential elements is reversed  
and the sign factor is calculated to be
 $ \sigma=\overline{n}+\frac{1}{2}(n+n')(n+n'-1)$.

We define a new Grassmann vector $z$ as 
$z=(\xi^*_{\nu'_1}, \cdots, \xi^*_{\nu'_{n'}}, \eta_{\nu_1}, \cdots, \eta_{\nu_{n}})$.
With a relation  Eq.(\ref{Grassmann_pf}) 
connecting the Pfaffian with the Grassmann integral, we obtain 
\begin{eqnarray}
\mathcal{M}_I&=&(-)^{\sigma}\int d z_{n+n'} \cdots dz_{1} 
e^{\frac{1}{2} \sum (\mathbb{M'}_I)_{\nu,\nu'}z_{\nu} z_{\nu'}} \nonumber \\
&=& (-)^{\sigma} {\it Pf}( \mathbb{M'}_I  ),
\label{Mabc4}
\end{eqnarray}
where $ \mathbb{M'}_I $ is a sub-matrix 
of the skew-symmetric matrix $ \mathbb{M'}$ 
 \begin{equation}
  \mathbb{M'} = \left(\begin{array}{cc}
      -B & C\\
      -C^T & -A\\
    \end{array}\right),
 \label{M}
 \end{equation}
concerning the index set $I$.
Block matrices $A$, $B$ and $C$ in $\mathbb{M}'$
are the contractions given in Eq.(\ref{ABC3}).
The dimensions of $\mathbb{M'}_I$ and $\mathbb{M'}$
are given to be $(n+n')$ and  $2M$, respectively.

The Pfaffian in Eq.(\ref{Mabc4}) can be transformed further as 
\begin{eqnarray}
 {\it Pf}(\mathbb{M'}_I)& =& {\it Pf} \left( \left(\begin{array}{cc}
    -B & C\\
    -C^T & -A\\
    \end{array}\right)_I \right) \nonumber \\
& =&(-)^{\frac{n+n'}{2}} {\it Pf} \left( \left(\begin{array}{cc}
       B & -C\\
       C^T & A\\
    \end{array}\right)_I \right) \nonumber \\
& =&(-)^{\frac{n+n'}{2}} {\it Pf} \left(
    \left( 
    K^T    
    \left(\begin{array}{cc}
       B &  C\\
       -C^T & A\\
    \end{array}\right)
    K
     \right)_I \right)  \nonumber \\
& =&(-)^{\frac{n+n'}{2}+n} {\it Pf} \left(    
    \left(\begin{array}{cc}
       B &  C\\
       -C^T & A\\
    \end{array}\right)
    _I \right),
\end{eqnarray}
where $
 K=   \left(\begin{array}{cc}
       1 &  0\\
       0 & -1\\
    \end{array}\right) 
$ and we use the Eq.(\ref{detpf}) in the last line.

Finally  we obtain a new and compact Pfaffian formula for  the matrix elements $\mathcal{M}_I$ as 
\begin{equation}
\mathcal{M}_I={\it Pf}( \mathbb{M}_I  ),
\label{Mabc5}
\end{equation}
where 
 \begin{equation}
  \mathbb{M} = \left(\begin{array}{cc}
      B & C\\
      -C^T & A\\
    \end{array}\right).
 \label{M_final}
 \end{equation}
Matrix  $\mathbb{M}$ has a bipartite structure, consisting of $A$, $B$ and $C$
in the Balian-Br\'ezin decomposition Eq.(\ref{abc}).
This structure is quite similar to that of the previous one obtained in
Ref. \cite{MO12}, but in the present formula, the definition of the
contraction is different due to the presence of the unitary operator $[\theta]$.
  
\section{Conclusion}
In this paper, we presented a compact Pfaffian
formula for matrix elements of a general unitary operator 
between  any multi quasi-particle HFB states.
To obtain this new Pfaffian formula, we use the Fermion coherent state and Grassmann integral.

This kind of matrix elements has been conventionally evaluated 
by means of the extended Wick's theorem
\cite{BB69}. 
The evaluation of matrix elements by the Wick theorem
was studied extensively for multi quasi-particle HFB states  
by Hara and Iwasaki in Ref. \cite{HI79}, which is the
essential component in the calculations of 
the Projected Shell Model (PSM) \cite{HS95}. 
The obtained result by Hara and Iwasaki, however, suffers from
the problem of combinatorial complexity arising from the Wick's theorem.
Standard PSM calculations for more than four quasi-particle HFB states 
are practically difficult.

In the present work, we were successful to find a general expression  
for the matrix elements in terms of the Pfaffian,
which has a more compact and closed form than the result obtained 
by Hara and Iwasaki.
Recently, Ref. \cite{BR12} shows a Pfaffian formula 
to matrix elements similar to Eq.(\ref{overlap}),
but our Pfaffian formula is physically intuitive due to the 
presence of the contractions.

The matrix in our formula has a bipartite structure and consists of the
matrices $A$, $B$ and $C$ appeared in the Balian-Br\'ezin decomposition.
This structure is quite similar to the previous one obtained in
Ref. \cite{MO12} where the matrix in the formula is expressed in 
a bipartite form of the skew-symmetric contraction matrix with 
density and pairing tensor.
The both formulae is found to be expressed in terms of the Pfaffian and
the generalized contraction matrices.

\section{acknowledgement}
Research at SJTU was supported by the National Natural Science
Foundation of China (Nos. 11075103 and 11135005), and by the 973
Program of China (No. 2013CB834401).

\section*{Appendix}
The Pfaffian is defined as   
\begin{equation}
\text{\it Pf}(A)\displaystyle \equiv\frac{1}{2^{n}n!}\sum_{\sigma\in S_{2n}}
{\rm sgn}(\sigma)\prod_{i=1}^{n}a_{\sigma(2i-1)\sigma(2i)}
\label{defpfaff}
\end{equation}
for a skew-symmetric matrix $A$ with dimension $2n\times 2n$, of which matrix elements are $a_{ij}$.
The $\sigma$ is a permutation of $\{1,2,3,\cdots , 2n\}$, ${\rm sgn}(\sigma)$ is its sign,
and $S_{2n}$ represents a symmetry group. For an $n \times n$ $(n=\text{odd})$ matrix,
$\text{\it Pf}(A) =0$. For a $2 \times 2$ matrix,  $\text{\it Pf}(A) =a_{12}$.
For a $4 \times 4$ matrix, $\text{\it Pf}(A) =a_{12}a_{34}-a_{13}a_{24}+a_{14}a_{23}$.

For a matrix   $P$ with dimension $2n \times 2n$, a following relation holds as
\begin{equation}
{\it Pf}(P^T A P)={\it det}Q {\it Pf} (A).
\label{detpf}
\end{equation}   

A relation between the Pfaffian and Grassmann integral is presented in Refs.\cite{MO12,ZJ02} as
\begin{equation}
 \int d\theta_{2n} \cdots d\theta_1
 \exp\left(\frac{1}{2}\theta^t A \theta\right)=\text{\it Pf}(A),
 \label{Grassmann_pf}
\end{equation}
where $\theta_1, \theta_2, \cdots, \theta_{2n}$ are Grassmann variables and 
$\theta^t=\left(\theta_1, \theta_2, \cdots, \theta_{2n}\right)$ is a Grassmann vector.
Matrix $A$ is a skew-symmetric matrix with $2n \times 2n$ dimension.

An overlap between two Fermion coherent states  $|\xi\rangle$ and $|\eta\rangle$ is given by \cite{NO85}
\begin{eqnarray}
\langle \xi| \eta \rangle &=& \langle \phi| e^{-\sum_{\nu}\xi^*_{\nu}a_{\nu}}e^{-\sum_{\nu}\eta_{\nu}a_{\nu}^{\dag}}|\phi \rangle \nonumber \\
 &=& \langle \phi| e^{-\sum_{\nu}\eta_{\nu}a_{\nu}^{\dag}}e^{-\sum_{\nu}\xi^*_{\nu}a_{\nu}}
 e^{ [-\sum_{\nu}\xi^*_{\nu}a_{\nu},-\sum_{\nu'}\eta_{\nu'}a_{\nu'}^{\dag} ]} |\phi \rangle \nonumber \\
 &=&  e^{ \sum_{\nu}\xi^*_{\nu} \eta_{\nu} }. 
 \label{grassmann-overlap}
\end{eqnarray}

Finally we prove a relation $e^{\sum (ln C)_{\nu,\nu'}a^{\dagger}_{\nu}
a_{\nu'}}|\xi\rangle=|C \xi\rangle$.
By the definition Eq.(\ref{FCS}) and $\hat{C}=\sum (ln C)_{\nu,\nu'}a^{\dagger}_{\nu}a_{\nu'}$, 
\begin{eqnarray}
e^{\hat{C}}|\xi\rangle&=&e^{\hat{C}} e^{-\sum_{\nu}\xi_{\nu}a_{\nu}^{\dag}}|\phi \rangle \nonumber \\
&=& e^{-\sum_{\nu}\xi_{\nu}e^{\hat{C}}a_{\nu}^{\dag}e^{-\hat{C}}}e^{\hat{C}}|\phi \rangle  \nonumber \\
&=& e^{-\sum_{\nu}\xi_{\nu} C_{\nu',\nu} a_{\nu'}^{\dag} }  \nonumber \\
&=& |\sum_{\nu}C_{\nu',\nu} \xi_{\nu}  \rangle 
\label{grassmann-a+a}
\end{eqnarray}
where $e^{\hat{C}}|\phi \rangle=|\phi \rangle$ and $e^{\hat{C}}a_{\nu}^{\dag}e^{-\hat{C}}=\sum C_{\nu',\nu}a^{\dag}_{\nu'}$.


\begin{thebibliography}{99}
  \bibitem{OY66} N. Onishi and S. Yoshida, Nucl. Phys. {\bf 80}, 367 (1966).
  \bibitem{HI79} K. Hara and S. Iwasaki, Nucl. Phys. A{\bf 332}, 61 (1979).
  \bibitem{NW83} K. Neerg\aa rd and E. W\"ust, Nucl. Phys. A{\bf 402}, 311 (1983).
  \bibitem{HHR80} K. Hara, Hayashi, P. Ring, Nucl. Phys. A385, 14 (1982)
  \bibitem{OT05} M. Oi and N. Tajima, Phys. Lett. B (2005).
  \bibitem{Rob09} L. M. Robledo, Phys. Rev. C {\bf 79}, 021302(R) (2009).  
  \bibitem{NO85} J. W. Negele and H. Orland, {\it Quantum Many-Particle Systems} (Westview Press, Oxford, 1998)
  \bibitem{BR12} G. F. Bertsch and L. M. Robledo, Phys. Rev. Lett. {\bf 108}, 042505 (2012).
  \bibitem{MO12} T. Mizusaki and M. Oi, Phys. Lett. B {\bf 715}, 219 (2012).  
  \bibitem{OM11} M. Oi and T. Mizusaki, Phys. Lett. B {\bf 707}, 305 (2012).
  \bibitem{AB12} B. Avez and M. Bender, Phys. Rev. C {\bf 85}, 034325 (2012).
  \bibitem{Rob07} S. Perez-Martin and L. M. Robledo, Phys. Rev. C 76, 064314 (2007).
  \bibitem{RS80} P. Ring and P. Schuck, {\it The Nuclear Many-Body Problem} (Springer, 1980).
  \bibitem{HS95} K. Hara and Y. Sun, Int. J. Mod. Phys. E {\bf 4}, 637 (1995). 
  \bibitem{BB69} R. Balian and E. Br\'ezin, Il Nuovo Cimento B {\bf 64}, 37 (1969).
  \bibitem{ZJ02} For instance, J. Zinn-Justin, {\it Quantum Field Theory and Critical Phenomena}
(Oxford University Press, London, 2002).

\end{thebibliography}
\end{document}